# PROPAGATION AND ENERGY DEPOSITION OF COSMIC RAYS' MUONS ON TERRESTRIAL ENVIRONMENTS

Franciole Marinho[1*], Laura Paulucci[2], Douglas Galante[3]

[1]Universidade Federal do Rio de Janeiro, Av. Aluizio Gomes, 50, 27930-560, Macaé, RJ, Brazil

[2]Universidade Federal do ABC, Rua Santa Adélia, 166, 09210-170, Santo André, SP, Brazil

[3]National Center of Research in Energy and Materials, Brazilian Synchrotron Light Laboratory, Campinas, Brazil

[*]Contact email: marinho@macae.ufrj.br

## Abstract

Earth is constantly struck by radiation coming from the interstellar medium. The very low energy end of the spectrum is shielded by the geomagnetic field but charged particles with energies higher than the geomagnetic cutoff will penetrate the atmosphere and are likely to interact, giving rise to secondary particles. Some astrophysical events, such as gamma ray bursts and supernovae, when happening at short distances, may affect the planet's biosphere due to the temporary enhanced radiation flux. Muons are abundantly produced by high energy cosmic rays in the Earth's atmosphere. These particles, due to their low cross section, are able to penetrate deep underground and underwater, with the possibility of affecting biological niches normally considered shielded from radiation. We investigate the interaction of muons produced by high energy cosmic rays on Earth's atmosphere using the Geant4 toolkit. We analyze penetration power in water and crust and also the interaction effects within bacteria-like material according to particle type and energy, and notice the possibility of off-track damage due to secondary particles.

1.   Introduction

Many astrophysical events can accelerate particles through shock waves and/or intense magnetic fields generating a cosmic ray flux that fills the interstellar medium. This cosmic ray flux has an energy spectrum extending for a wide range from about $10^3$ eV up to $10^{19-20}$ eV (see, for example Yao et al., 2006) and it has been shown by Badhwar and collaborators (Badhwar et al., 1994) that charged particles are one of the most damaging physical phenomena in space environment.

The geomagnetic field protects Earth from the most intense flux of low energy particles and although particles penetrating the atmosphere can have energies of up to $10^{20}$ eV, a very strong suppression of the ultra-high-energy cosmic rays flux is observed (Abraham et al., 2010). Some astrophysical events may temporarily enhance the high energy flux of radiation. For instance, gamma-ray bursts (Thorsett 1995, Atri et al. 2013), supernovae (Terry & Tucker 1968), and very large solar flares (Dennis 1985) emit intense high energy radiation (photons and charged particles) that could cause important biological effects.

Ionizing radiation can cause cellular damage both by direct breaking of the DNA chain (single and double strand breaks) or by creation of free radicals. The hydroxil radical (HO$^\bullet$) that can form through water molecule ionization is a strong oxidant that can interact with many different kinds of molecules within the cell. Besides the hydroxil radical, hydride radicals (H$^\bullet$) and electrons removed in the ionization process can interact with macro-molecules, such as proteins, lipids, and DNA itself, disrupting them. When suffering this kind of interaction, the cell can either die or regenerate. This regeneration may be imperfect leading to mutations with possible consequence of a runway multiplication with serious biological implications for a multicellular organism or a colony of unicellular individuals.

The usual approach for analyzing the interaction of particle radiation with living cells is experimental, meaning that irradiation assays are performed in order to obtain a relation between survivability and radiation flux. This methodology although reproducible, gives little information on the physical mechanisms taking place with different particles and energies. In addition, it is difficult to extrapolate the data in order to predict the survivability under conditions which cannot be experimentally replicated. This is exactly the case when studying the influence of a muon flux on the biosphere. These particles are produced in Earth's atmosphere by interaction of cosmic rays and although several other types of particles are also produced, muons are the dominant component of the radiation flux at sea level for energies above 100 MeV (Alpen, 1998). The effect of the muons is complementary to that of the hadronic and other electromagnetic components, that is electrons, nuclei, and photons, which are more efficient in depositing energy on the exposed biota (Dar et al., 1998; Thomas et al., 2005; Juckett, 2009; Karam, 1999, 2002a,b).

However, since muons are low interacting particles, they can reach great depths before causing substantial energy deposition, making it a possible vector of undersea and underground biological damage. Moreover, their effects on living organisms are difficult to be directly measured. In order to have a better understanding of the physical processes caused by muon interaction in Earth environments of biological relevance we have employed the Geant4 simulation toolkit (Agostinelli et al., 2003) to track muons and all secondary particles produced and to obtain the energy deposited along their paths.

2. Production of muons

Muons can be produced by the interaction of charged particles (*A*) with atoms on the planet's upper atmosphere, represented on equation 1 by *N*. These interactions produce pions ($\pi$) and kaons (*K*) which decay into muons ($\mu$) and neutrinos ($\nu$) that can propagate deep through the crust and water due to their low interaction cross section.

$$A + N \rightarrow \pi + K \rightarrow \mu + \nu \qquad (1)$$

The flux of muons at sea level was simulated numerically starting from the interaction of a proton of 10 TeV of energy interacting in the top of the atmosphere at normal incidence. The first interacting particle is often referred to as the primary particle. Earth's atmosphere was based on the NRL-MSISE-00 model (Picone et al., 2002) which presents profiles for the elements density as a function of altitude considering molecules proportions currently observed. The geometry of the atmosphere was described as a sequence of spherical shells with varying widths and assuming average density and composition. The absolute difference between density values at the top and bottom of each shell was required to be smaller than a few percent.

The Geant4 package is built in order to best simulate interactions using both theoretical models and up-to-date data, when available. It considers muons interactions for both continuous and discrete energy losses, including ionization, bremsstrahlung, positron-electron pair production and muon photonuclear interaction with nuclei. Comparing the muon flux obtained with Geant4 as depicted in fig. 1 with the one calculated by Atri & Mellot (Atri and Mellot, 2011) using the software CORSIKA, we see similar results for the energy spectrum. We note that for off-zenith incidence the spectrum would have its peak shifted towards lower energies.

Muons can also be produced in the atmosphere by photonuclear interactions. Photons producing such interactions must be very energetic (at least ~10 GeV) and a copious number of muons could be generated this way by the influence of a nearby short gamma ray burst (Atri et al. 2013). Although we did not consider this reaction for the generation of muons, results presented here can be useful when investigating such scenario.

## 2.1    Muons at sea level and underground

A substantial mass of the terrestrial microbiota is on radiation-shielded environments, such as underwater or underground. If a high-energy astrophysical event happened in the vicinities of Earth, it would be interesting to evaluate the possible damage over this apparently protected niches caused by particles that could penetrate the barrier provided by the geomagnetic field and that would not be effectively stopped by the thick atmosphere layer.

We have simulated a biological material with density of 1.1g/cm$^3$ and composition of 31% C, 49% H, 13% O and 7% N. The elemental composition used was a general average for most bacteria, which is described in further detail in the works of Porter (1946), Salton (1964), and Hiragi (1972), but more modern data is scarce. These values are reasonably accurate for the most abundant elements, but highly variable for those forming just a few percent of the material. Although, these small fractions of components may produce large differences on the response of microorganisms in face of biological radiation damage they do not significantly affect the estimated energy deposition mechanisms. For instance, *Deinoccocus radiodurans*, a highly radiation resistant organism, has a larger intracellular concentration of Mn and lower concentration of Fe, which was described as being correlated to its resistance against ionizing radiation (Daly et al, 2004, Daly, 2009).

For the crust we used a material with density 2.6 g/cm$^3$ composed mainly of 46.6% O and 27.7% Si and lower quantities of 8.1% Al, 5.0% Fe, 3.6% Ca, 2.8% Na, 2.6% K, 2.1% Mg, 0.5% Ti, and 0.2% H.

We show in fig. 1 the energy spectra of muons considering penetration in crust and water compared to the one expected at sea level. We see that the high energy end of the spectrum is much more attenuated in crust than in water, where the low energy region is more substantially affected.

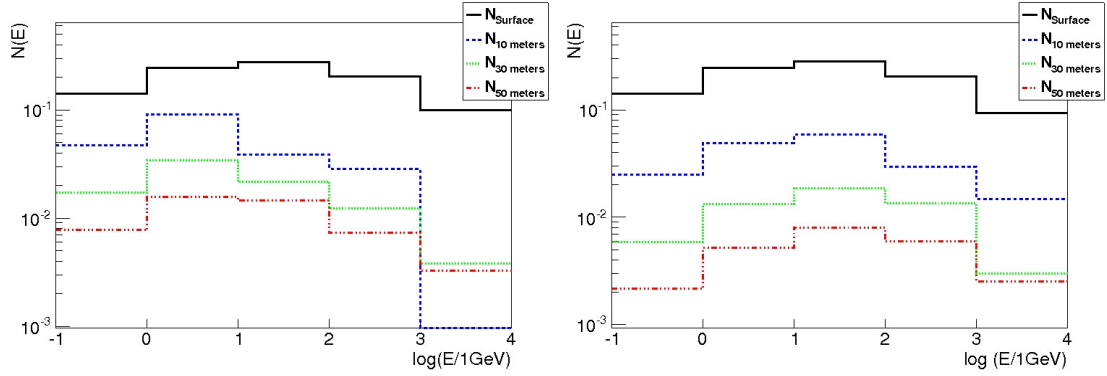

Fig 1: Left panel: Muon spectrum on Earth's surface and underground, produced by 10TeV proton primaries on the top of the atmosphere. Right panel: Same calculations under water.

In fig. 2, we show the fraction of muons relative to the sea level flux that can reach a certain depth in crust. The flux is reduced by a factor of ~10 in the first 100 meters. After ~3.5 km below the surface, the quantity of muons is greatly reduced, though until this depth it shows a milder reduction with distance traveled.

This behavior is closely related to the muon stopping power. The stopping power ($-\langle dE/dx \rangle$) is the rate at which a charged particle loses energy per distance unit traveled. Since there are many possible interactions for a given particle and the probabilities associated to them are energy dependent, so is the behavior of the stopping power. For the case of muons, the stopping power presents a minimum (minimum ionizing particle) for mid energies (depending on the material but typically in the ~GeV region) whereas the rate of energy loss is greatly enhanced at the low (~MeV) and high energy ends (~TeV) (Beringer et al., 2012). It leads to the behavior seen in Fig. 2: low energy muons in the initial flux are easily stopped within the crust, causing the first drop in the curve; high energy muons also quickly loose energy increasing the number of mid energy muons, which can travel much longer within the crust because of the smaller stopping power, leading to the milder reduction seen in the spectrum. They continue loosing energy at a small rate until their energy is such that the stopping power increases again (low energy range) and they quickly loose energy again leading to the fast decay in the spectrum seen at greater depths.

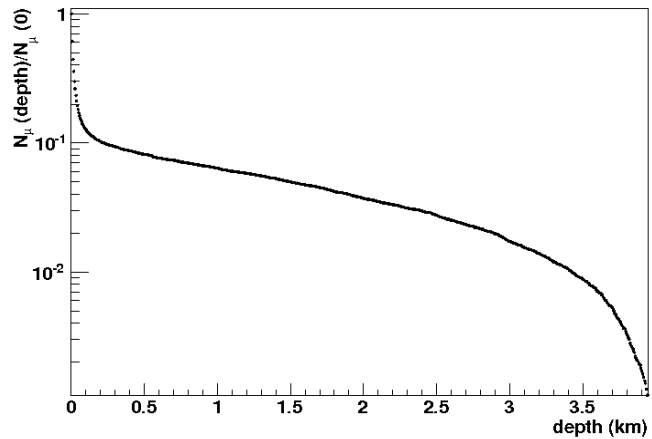

Fig 2: Fraction of muons reaching a given depth underground with respect to the sea level flux.

## 2.2 Energy Deposition and Mean Free Path

We have calculated the deposition of energy by muons in water, crust and also considering a material that could fairly simulate a large colony of bacteria in a close packing formation. The results are shown in fig. 3 for a 1 GeV muon with orthogonal incidence in matter and considering all possible mechanisms of interaction. The Bragg peak is evident in all distributions, being that in the crust the peak is closer to the surface while in water, it is farther away, providing less shielding. This peak structure is usually seen when heavy charged particles increase their energy deposition in the media as they slow down. Therefore, a sudden increase in the energy deposition is concentrated in a small region leading to a peak at the end of the particle trajectory.

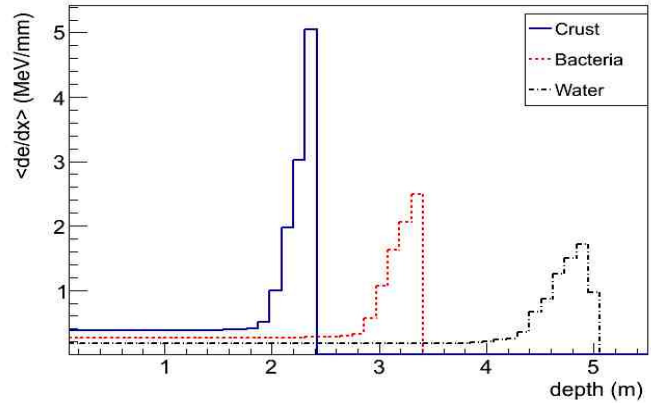

Fig 3: Energy deposition curves on water, crust and bacteria-like material for 1GeV muons with normal incidence angle.

The muon mean free path (MFP) was also obtained on different types of media (water, crust and bacteria-like material). On fig. 4, the MFP for the bacteria-like material shows that 1GeV muons would be able to travel for large distances after a first hit, much larger than a cell radius (typically of the order of 1 μm), on the linear energy deposition region, unless the muons lose sufficient energy on the substrate to be on the Bragg curve regime.

The number of DNA double strand breaks (DSB) for a given gamma radiation dose is ~0.004 DSB/Gy/Genome for the *Deinococcus radiodurans* (see Daly et al. 2004 and references therein). This value is very similar to other type of non-radiation-resistant bacteria. The relative biological effectiveness of muons is similar to that of gammas (Atri et al., 2013). For muons, Fig. 3 and 4 show that the number of DSB would be on average 0.0025 DSB/Gy/Genome.

A first muon hit typically produces one secondary electron, for which the MFP is also shown on fig. 4. This electron would travel for a few mm before producing a hit and possible secondary molecular damage, creating tracks around the muon primary path on the matter. Low energy electrons are also produced close to the muon path having MFP estimated to be 0.1 mm or lower. The uncertainties on the MFP values calculated on the right panel are proportional to the inverse square root of the number of electrons per event. Given the energy conservation principle, the total production of electrons with higher energies is less probable than the ones with lower energies, being the maximum energy of a single electron limited by the energy of the incoming muon. In this way, the counting for the low energy range will be greater, rendering smaller error bars.

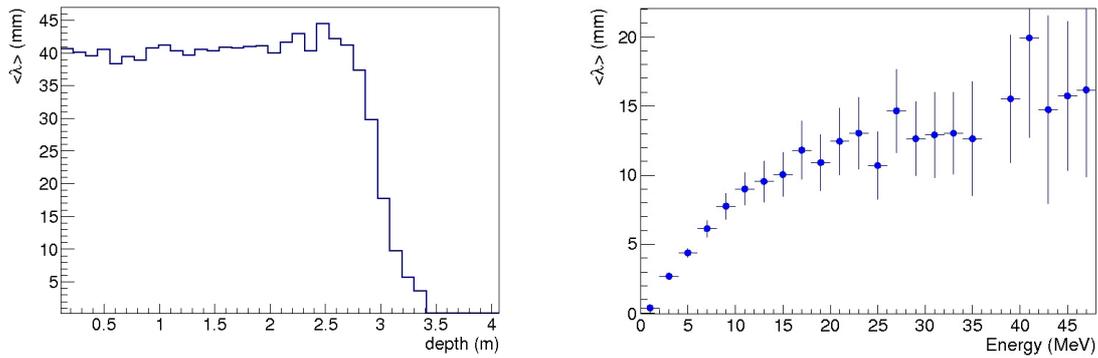

Fig 4: Mean free path for 1 GeV muons at normal incidence angle (left) and electrons produced by them (right) in a material emulating a bacteria colony.

## 2.3    Generation of Secondary Particles

The main reaction caused by muon interactions in the deepest range achieved is ionization, giving rise to several electrons and low energy photons that will propagate and further interact. These are potentially dangerous for they can produce free radicals within cells. The necessary energy for ionizing a water molecule, which comprises about 80% of a cell, ranges from ~13 to 19 eV and is much lower than the typical energies deposited (in the keV range) along these particles path.

It is also important to note that muons with energies higher than 1 GeV will penetrate deeper in the substrate, either water, crust or the material emulating a bacteria community. It means that they leave a trace of secondary electrons along a much longer path, potentially damaging more members of a microbial community and this damage not being restricted to surface layers.

The muon lateral displacement seen in fig. 5 indicates that, besides not being shielded by crust or water at moderate depths, the microbial community could suffer damage over an extensive area.

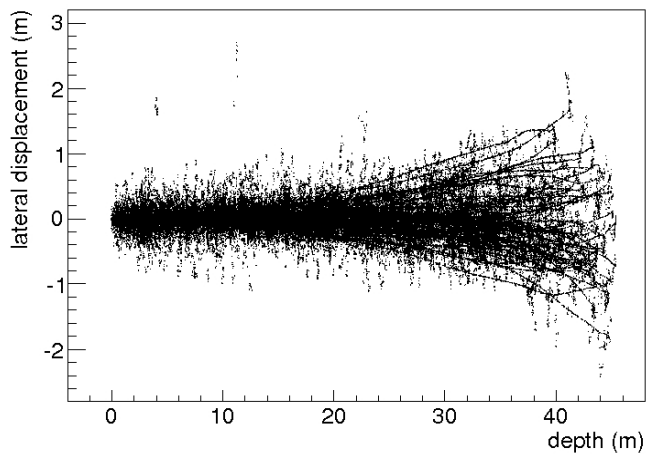

Figure 5: Interaction points of a 10GeV muon trajectory with electron production.

The electrons and positrons generated by the muon will have energies above 100 keV. These particles will later interact throughout the material ionizing it and releasing lower energy electrons also close to the muon trajectory.

The expected energy distribution of photons produced along the muon path is illustrated in figure 6. These are also a source of a cascade production of a copious number of low energy electrons with MFP of about tenths to hundreds of nanometers via processes such as Compton scattering and photoelectric effect. These particles could potentially damage bacteria cells since they would generate microscopic electronic clouds around their paths.

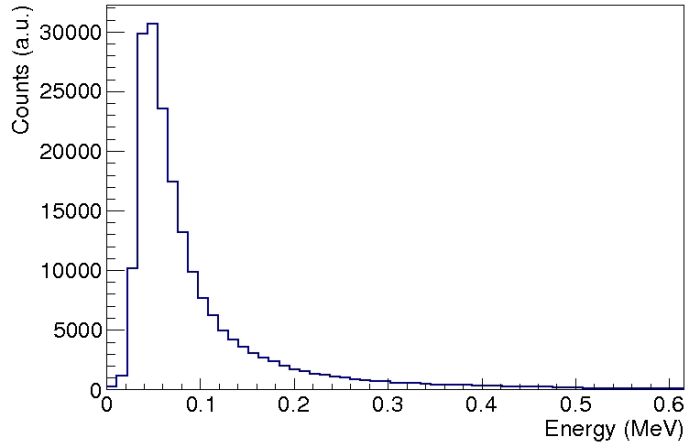

Figure 6: Typical photon energy distribution produced by a 1GeV muon.

3. Conclusions

We present the simulation of a muon flux caused by a primary proton interaction on the atmosphere and propagated onto the surface of the Earth. These particles were propagated through matter by the Geant4 toolkit, indicating that still at 4km of depth (crust) a significant fraction of the initial muon flux ($10^{-3}$) would be present, due to the low probability of interaction of these particles, possibly reaching biological niches otherwise protected.

We have analyzed the interactions in a bacteria-like material designed to emulate a microbial community in close packing formation as a function of particle type and energy, as well as a function of its surrounding conditions, i.e., the presence of materials that could either block, slow down or produce secondary radiation of biological importance.

Muons can potentially produce damages on molecules depending on the flux, such as direct strand breaks on DNA, and ionizations. The secondary electrons produced in these processes can also create molecular breaks on a distance range of a few millimeters off the track of the muons, creating a typical sidetrack effect usually seen on laboratory experiments or cosmic ray exposure. Recent estimates (Daly 2009) show that the probability of double strand breaks is not alone responsible for immediate death of an individual since the difference is minimal between these values for radiation resistant bacterias such as the *Deinococcus radiodurans* and less resistant bacteria. It seems that damage in enzymes that repair DNA caused by the passage of ionizing radiation plays a fundamental role. Damage in these repairing enzymes are mainly caused by oxidant stress due to the presence of free radicals within cell.

This study provides relevant information for estimating the resilience of life on environments subjected to high muon fluxes, which can be exposed surfaces of planets, with or without an atmosphere, or Earth itself, during an event of elevated bombardment by cosmic rays, such as on the presence of a nearby supernova or even the weakening or inversion of the planetary magnetic field.

Acknowledgements

The authors would like to thank the Brazilian funding agencies FAPERJ, CNPq and FAPESP for financial support.

# References

Abraham, A.J., et al. (The Pierre Auger Collaboration), 2010. Phys. Lett. B, 685, pp. 239–246.

Agostinelli, S., Allison, J., Amako, K., et al., 2003. Nucl. Instrum. Meth. A, 506(3), pp. 250–303.

Alpen, E.L., 1998. Radiation Biophysics. Academic Press, San Diego.

Atri, D., Mellot, A.L., 2011. Radiat. Phys. Chem. 80, pp. 701.

Atri, D., Mellot, A.L., Karam, A., 2013, International Journal of Astrobiology, FirstView Article, pp. 1.

Badhwar, G.D., Cucinotta, F.A., Oneill, P.M., 1994. Radiat. Res., 138(2), pp. 201–208.

Beringer, J., et al., 2012. Phys. Rev. D, 86, pp. 010001.

Daly, M.J., et al., 2004, Science, 306, pp. 1025-1028 .

Daly, M.J., 2009, Nature Reviews Microbiology, 7, pp. 237-245.

Dar, A., Laor, A., Shaviv, N.J., 1998. Phys. Rev. Lett., 80(26), pp. 5813–5816.

Hiragi, Y., 1972, J. Gen. Microbiol., 72, 87.

Juckett, D.A., 2009. Int. J. Biometerol., 53, pp. 6.

Karam, P.A., 1999. Health Phys., 77, pp. 6.

Karam, P.A., 2002a. Health Phys., 82, pp. 4.

Karam, P.A., 2002b. Radiat. Phys. Chem., 64, pp. 2.

Picone, J.M., Hedin, A.E., Drob, D.P., Aikin, A.C., 2002. J. Geophys. Res., 107, SIA 15.

Porter, J.R., 1946, Bacterial Chemistry and Physiology. London: John Wiley and Sons.

Salton, M.R.J., 1964, The Bacterial Cell Wall, Elsevier, Amsterdam.

Terry, K.D., Tucker, W.H., 1968, Science 159 (3813), pp. 421.

Thomas, B.C., et al., 2005. Astrophys. J., 634, pp. 509.

Thorsett, S.E., 1995, Astrophys. J., 444, pp. L53.

Yao, M.W., et al., 2006. J. Phys. G, 33, pp. 245.